\documentclass[aps,pra,twocolumn,final,floatfix,superscriptaddress,10pt]{revtex4}
\usepackage{amsthm}
\usepackage{latexsym}
\usepackage{graphicx}
\usepackage{hyperref}
\usepackage{natbib}
\usepackage{amssymb}
\usepackage{nicefrac}
\usepackage[utf8]{inputenc}
\usepackage{dsfont}
\usepackage{xcolor}
\usepackage{tikz}
\usepackage{nicefrac}
\usepackage{ulem}
\usepackage{qcircuit}
\usepackage{dcolumn}   
\usepackage{mathtools}
\usepackage{braket}
\usepackage{soul}
\usepackage{cancel}
\usepackage{mathdots}
\usepackage{subfigure}
\usepackage{placeins}
\usepackage{lipsum}
\usepackage{float}

\usetikzlibrary{arrows,shapes,decorations.pathmorphing}

\makeindex

\begin{document}

\title{Linear and logarithmic time compositions of quantum many-body operators}
\author{F. Motzoi}
\affiliation{Department of Physics and Astronomy, 8000 Aarhus, 	Denmark}
\author{M. P. Kaicher}
\affiliation{Theoretical Physics, Saarland University, 66123
	Saarbr{\"u}cken, Germany}
\author{F. K. Wilhelm}
\affiliation{Theoretical Physics, Saarland University, 66123
	Saarbr{\"u}cken, Germany}

\begin{abstract}
We develop a generalized framework for constructing many-body-interaction operations either in linear time, or in logarithmic time with a linear number of ancilla qubits. Exact gate decompositions are given in particular for Pauli strings, many-control Toffoli gates, number-~and parity-conserving interactions, Unitary Coupled Cluster operations, and sparse matrix generators. We provide a linear time protocol that works by creating a superposition of exponentially many different possible operator strings and then uses dynamical decoupling methodology to undo all the unwanted terms. A logarithmic time protocol overcomes the speed limit of the first by using ancilla registers to condition evolution to the support of the desired many-body interaction before using parallel chaining operations to expand the string length.  The two techniques improve substantially on current strategies (reductions in time and space can range from linear to exponential), are applicable to different physical interaction mechanisms such as CNOT, $XX$, and $XX+YY$, and generalize to a wide range of many-body operators.
\end{abstract}

\maketitle

\section{Introduction}
Generating multi-body entanglement is the hallmark of most quantum information technologies.  Such technologies promise to harness entanglement across multiple quantum registers to enable potentially significant improvements in speed or precision compared to their classical counterparts. Yet much of the difficulty in the control of quantum systems lies in the constraint that entanglement naturally arises on a local scale while scaling improvements occur as a result of wave functions spread over much larger spaces. 

Quantum circuits generating entanglement across $n$ qubits in linear or sub-linear time (circuit depth) in $n$ have been the subject of many studies, with direct use as subroutines in quantum algorithms for factoring \cite{cleve2000fast, shor}, simulation
\cite{lloyd0,harrow2009quantum,aag1,barends1,aag2,aag3,troyer,ryan1,wiebe2011simulating,jarrod1,pl,borzu,ryan2}, unstructured search \cite{molmer2011efficient, page}, error-correction \cite{crow2016improved}, and solutions to systems of differential equations \cite{harrow2009quantum,berry,aharonov2003adiabatic,wiebe2011simulating}.  Much progress has been made for constructing many-body operations, with the most success coming from either finding architectures where commuting two-qubit interactions could be executed simultaneously on overlapping Hilbert spaces, or via cases where a particular many-body gate with known or suspected sublinear implementation can be used to synthesize other many-body circuits.  The former has been used for so-called \emph{collective} Pauli operations on qubits in ion chains \cite{molmer,blatt2000quantum,monz201114}, while the workhorse for the latter has been the \emph{fanout} operation\cite{moore2001parallel,fenner2005bounds}, which has successfully lead to $O(\log(n))$ depth quantum circuits for various flavours of quantum adders \cite{takahashi2008fast,gossett1998quantum,svore}, with related arithmetic operations \cite{van2005fast,cleve2000fast}.  

Other many-body interactions have also been synthesized to mixed success.  The ubiquitous many-control CNOT has found general linear-depth implementations, though with a relatively large prefactor \cite{maslov1,barenco1,maslov}.  Another commonly used gate is defined in terms of the rotation between two arbitrary multi-qubit states, for use in sparse matrix generation \cite{aharonov2003adiabatic,wiebe2011simulating,berry2014exponential,harrow2009quantum}, or, equivalently, pairwise inversion of opposing spins in Unitary Coupled Cluster (UCC) theory \cite{bartlett1,bartlett2,jonathan,kutzelnigg,hoffmann}.  Here, suggested implementations have typically involved (based on intended application) either a linear ancilla memory operated on with multi-control rotations \cite{aharonov2003adiabatic,wiebe2011simulating}, or Trotter decomposition of the dynamics into (exponentially many) Pauli-string factors but with no ancillas.

In this work, we provide a generic formalism for how to directly compose a wide class of such many-body entangling operations (generated each by an equivalent Hermitian many-body operator $\bar H$)  via two-(or few-)local interactions, and for which the above discussed protocols and algorithms form important examples of its application.  We aim to minimize two standard figures of merit of the generic circuit, namely its \textit{depth}, defined as the number of layers of gates acting simultaneously on disjoint sets of qubits, and its \textit{width}, defined as the total number of qubits acted on by the circuit \cite{bera}. We find a width-optimized general algorithm, which we label the \textit{decoupling protocol}, to compose $\bar H$ with zero or constant memory overhead and depth limited to linear $n$ scaling. Moreover, we demonstrate a depth-optimized algorithm for simulating $\bar H$, the \textit{selection protocol}, which has logarithmic depth and requires at most linear memory overhead. We demonstrate the formalism towards a linear to exponential speed-up of the aforementioned examples, given in our notation by equivalent Hamiltonians
\begin{align*}
\bar H=\begin{cases}
& \prod_{i=1}^nX_i \ \quad\text{(n-qubit Pauli strings)}\\
&(\prod_{i=1}^n P_i)X_{n+1} \ \text{(n-controlled X gate)}\\
&\prod_{i=1}^{n/2}(\sigma^+_{2i-1}\sigma^-_{2i}+h.c.) \ \text{(number/parity consv.)}\\
&(\prod_{i=1}^n\sigma_i^+)+h.c.\  \text{(UCC-type operators,}\\
&\qquad \qquad\qquad\qquad\ \text{sparse matrix generators)},
\end{cases}
\end{align*}
where $X_i$,$Y_i$ and $Z_i$ are Pauli operators acting on qubit $i$, $\sigma_j^\pm=(X_j\mp iY_j)/2$ and projector $P_i={\ket{1}}\bra{1}_i$. The unitary circuit generated by the many-body composed dynamics can then be written succinctly via the notation
\begin{align}
[\bar H]^\alpha\equiv\exp(-i\alpha\bar H), 
\end{align}
with $\alpha$ the rotation angle. 


In order to identify the type of Hamiltonians we can compose, we first introduce notation and conventions. For simplicity, we disregard local unitary transformations between operators of the same rank. Let $\mathcal H=\mathcal H_1\otimes\mathcal H_2\otimes \dots \otimes \mathcal H_n$ denote a separable Hilbert space. Then $R_i=\mathds 1^{\otimes(i-1)}\otimes R\otimes \mathds 1^{\otimes(n-i)}$, where $\mathds 1_j$ denotes the identity on $\mathcal H_j$ and $R$ ($\neq \mathds 1$) is a $2\times 2$-Hermitian matrix for qubits (or $d\times d$ for qudits). 
We use the convention $\text{rank}(R_i)\equiv\text{rank}(R)$, thereby ignoring contributions from identities $\bigotimes_{\jmath\neq i}\mathds 1_j$ on other sub-Hilbert spaces.  
We can write a higher-rank $R_i$ as the tensor sum 
$R=V(S\oplus{S^\perp})V^\dagger$, where $V$ is any local unitary transformation 
 (i.e.~$S^\perp$ lies in the kernel of $S$). 
It is by chaining together the lower-rank $S_i$ factors that we will be able to construct our many-body dynamics $\bar H$.


\section{Decoupling protocol} 
\begin{figure}[h!]
	\includegraphics[width=.46\textwidth]{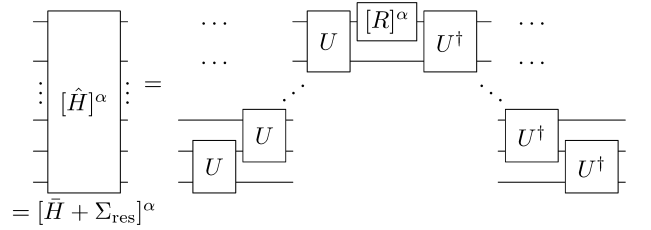}
	\caption{Generating the unitary dynamics $[\hat H]^\alpha=[\bar H+\Sigma_{\text{res}}]^\alpha$ of Eq.~(\ref{h}) using one single-body and $2n$ $U_{j-1,j}$ operators. \label{fig:0b}}
\end{figure}

\begin{figure}[h!]
	\includegraphics[width=.4\textwidth]{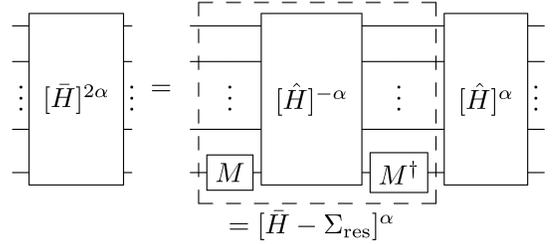}
	\caption{Gate sequence for realizing the decoupling protocol given in Eq. (\ref{dd}) using $4n$ $U_{j-1,j}$ operators.\label{fig:0a}}
\end{figure}

Our main tool is a unitary (two- or few-body) operator $U_{j-1,j}$ which will be used to iteratively increase in length a string of Hermitian operators $S_1S_2\cdots S_j$ acting on the system. However recall that generally $S_j$ will not be full rank, and so $U_{j-1,j}$ will invariably have to also act outside the support of $S$. Thus, our protocol will have to execute the desired system dynamics (given by $\bar H$) while leaving the rest of the Hilbert space (namely the kernels of $S_j$) intact.
A condition to using the protocol is that a $U_{i,j}$ can be found  such that
\begin{align}
U_{i,j}^\dag R_iU_{i,j}=S_iR_j+S_i^\perp+S_iR_j^\perp,  \label{u}
\end{align}
thereby incrementing the length of a string of non-identity Hermitian operators by one when acting on $R_i$. Then, successively applying Eq.~(\ref{u}), one can show that the following sum of operator strings of increasing string length can be composed 
\begin{align}
\hat H =& \left(\prod_{j=n}^1U_{j,j+1}^\dag\right)R_1\left(\prod_{j=1}^nU_{j,j+1}\right)\nonumber\\
=&\left(\prod_{i=1}^nS_i\right)R_{n+1}+\sum_{m=1}^n\left(\prod_{i=1}^{m-1}S_i\right)\left(S_{m}R_{m+1}^\perp+S_m^\perp\right)\nonumber\\
\equiv&\bar H+\Sigma_{\text{res}}, \label{h} 
\end{align}
as shown in Fig. \ref{fig:0b}, where $\bar H=(\prod_{i=1}^nS_i)R_{n+1}$. Therefore, composing $\bar H$ from two-body operators $U_{i,j}$ usually creates unwanted remainder terms $\Sigma_{\text{res}}$. However, the remainder terms commute with $\bar H$, while also acting as the identity on the support of  $R_{n+1}$.  Thus, we can find a one-body unitary transformation $M_{n+1}$ such that it imparts an opposite phase to $R_{n+1}$ (and thus $\bar H$), but does not  change $\Sigma_{\text{res}}$ \cite{mikecomment}. The requisite dynamics can then be recovered using the decoupling sequence (cf.~Fig.~\ref{fig:0a})
\begin{align}
[\bar H]^{2\alpha}=\left[ \hat H\right]^\alpha M_{n+1}^\dag \left[\hat H\right]^{-\alpha} M_{n+1}\label{dd}.
\end{align}
Here, to construct an effective Hamiltonian $\bar{H}$ of string-length $n+1$, a total of $4n$ unitary operators $U_{i,j}$ are needed. Note that if $R$ is full rank, Eq.~(\ref{u}) reduces to $U_{i,j}^\dag R_iU_{i,j}=R_iR_	j$ and only half as many operators are used, since decoupling is not required. 


\section{Selection Protocol}
\begin{figure}
	\includegraphics[width=.48\textwidth]{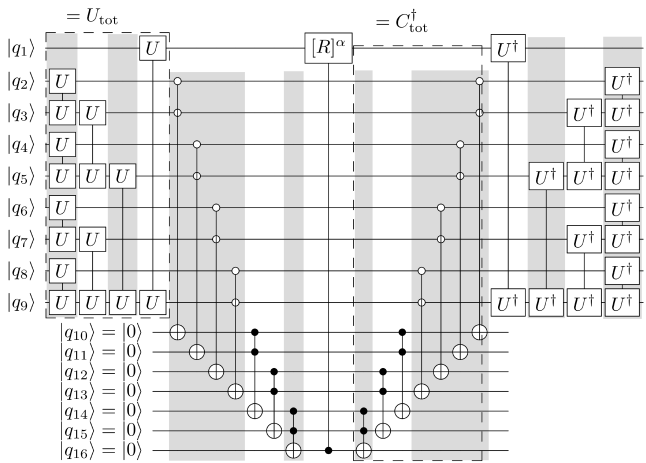}
	\caption{Gate sequence to compose an operator string of length $n=9$, to run in $O(\log n)$ depth. Here, we assume that $1<\text{rank}(S_i)<\text{rank}(S_i+S_i^\perp)$ for $i>1$. Small empty circles are conditioning operators (projectors) on the support of $S$ while solid circles are conventional Toffoli gates.  The connected $U$ are the entangling operations from Eq. (\ref{u}). Since $R_1\equiv S_1$ is only applied on the support of $S_2...S_9$, $\hat H$ (which is created by the $U_{i,j}$ tree structure) is also only applied on the support of $\bar H$. \label{fig:1}}
\end{figure}

\begin{figure}
	\includegraphics[width=.4\textwidth]{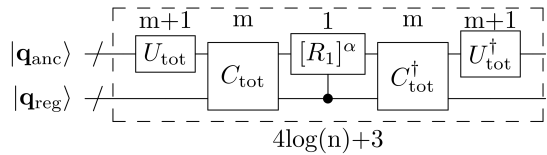}
	\caption{Generalized version of Fig. (\ref{fig:1}), where all register (resp. ancillary) qubits $\mathbf q_{\text{reg}}$ and $\mathbf q_{\text{anc}}$ are compounded to one circuit line. All operations that can be run in parallel are synthesized to one box, with the number above it indicating how many non-commutative time steps are necessary for each box. For $n=2^m$, a string of length $n+1$ is composed in $\mathcal{O}(\text{log}(n))$ depth. \label{fig:2}}
\end{figure}

We now present a composition scheme that can further decrease the required circuit depth for $n$-body operators from a $\mathcal{O}(n)$ to a $\mathcal{O}(\text{log}(n))$ scaling, at the cost of $n-1$ ancillas (cf. Figs. \ref{fig:1} and \ref{fig:2}).  Without loss of generality we set $n=2^m$ where $m\in\mathds{N}$ and introduce two sets of qubit indices, namely register qubits $q_{\text{reg}}=\{1,2,...,n+1\}$ containing the qubits of the desired string and ancilla qubits $q_{\text{anc}}=\{n+2,2,...,2n\}$, where the latter are all initialized to the $\ket{0}$ state. 

Moreover, we introduce Toffoli-type unitary operators $C_{i,j}X_{k}$, whereby a NOT is applied to qubit $k$ conditioned on the state of qubits $i,j$ which can either be register or ancilla qubits. The operation can be written mathematically as e.g.~$C_{i,j}X_{k}=[S_iS_jX_k]^{\pi/2}$. For $i,j\in q_{\text{reg}}$, $C_{i,j}$ essentially conditions on being in the support of both $S_i$ and $S_j$. If $i,j\in q_{\text{anc}}$, the operation is a standard Toffoli (or simply letting projector $S_i=\ket{1}\bra{1}_i$). Note that if $n$ is not a power of $2$, some $C_{i,j}X_k$ operations can have both register and ancilla qubits as controls.
We define as in Fig.~\ref{fig:1} compound operations
\begin{align}
C_{\text{tot}}\equiv&\prod_{k=1}^{\log n}\left(\prod_{l=n-2^{k}+1}^{n-2^{k-1}}C_{2l+1,2l}X_{n+l+1}\right),\\
U_{\text{tot}}\equiv&\prod^{1}_{k=\log n}\left(\prod_{l=1}^{n/2^{k}}U_{l 2^{k}+1,(l-\frac{1}{2}) 2^{k}+1} \right)U_{1,n+1}.\label{ctot}.
\end{align}
For the boundary case $\text{rank}(S)=1$, we set $U_{\text{tot}}=\mathds{1}$, while if $S$ full rank then $C_{\text{tot}}=\mathds{1}$.
All the operations in the brackets can be run parallel.
The full selection protocol is given by the sequence
\begin{align}
[\bar H]^\alpha=
U_{\text{tot}}^\dag C_{\text{tot}}^\dag\left(C[R_{1}]^\alpha\right)C_{\text{tot}} U_{\text{tot}}, \label{vc}
\end{align}
see Fig. \ref{fig:2}. The middle operator is defined as $C[R_{1}]^\alpha=[R_{1}]^\alpha$ if S full rank, and $[\ket{1}\bra{1}_{2n}R_{1}]^\alpha$ otherwise. The result of the sequence is that the many-body rotation is applied only on states that are supported by $R_{1}S_2 \cdots S_n$, while identity is applied otherwise.  The selection protocol improves on generic parallelization algorithms by quadratically reducing the space requirements \cite{moore2001parallel}.

The following sections demonstrate how to apply the \textit{decoupling} and \textit{selection protocols} to well-known existing problems, to either take advantage of specific two-body interaction mechanisms, or to reduce the time and gate complexity of known implementations. Further details of the derivations are given in Appendix B.

\section{Pauli strings}
Well known formulas exist for forming strings of Pauli operators in linear time (e.g. \cite{nielsen,mezzacapo}), which we first reproduce using our formalism. Since Pauli operators are full rank, $\hat H=S_1 S_2 \cdots S_j=\bar H$ and there is no need for decoupling. The optimal form of $U_{i,j}$ will depend on the architecture and its natural interaction. A standard entangling operation is  via $U_{i,j}=\text{CNOT}_{ij}$ gates \cite{nielsen}, which applied as in Fig.~\ref{fig:0b} produce a many-body operator \eqref{h} with $S_i=X_i$. However this interaction is neither natural for superconducting nor trapped ion qubits. A native  gate for ion-trap designs is the M\o{}lmer-S\o{}rensen gate (MSG), where commuting $U_{i,j}=[X_iX_j]^{\tfrac{\pi}{4}}$ \cite{molmer,molmerII} interactions can be applied simultaneously to all pairs of ions in the chain. We present a third composition with the same gate count as the CNOT and MSG circuits, designed for architectures with exchange gate interactions, $G_{i,j}=\frac{1}{2}(X_iX_j+Y_iY_j)$. This is the fastest perfect entangler for most circuit-QED quantum processors \cite{goerz, siewert}, as well as for quantum dot spins coupled by a cavity \cite{imamoglo} and nuclear spins interacting via a two-dimensional electron gas \cite{mozyrsky}. Here, $U_{i,j}=[G_{i,j}]^{-\tfrac{\pi}{2}}$ is an iSWAP gate. Using Eq. (\ref{u}) and $U_{i,j}^\dag Y_iU_{i,j}=Z_iX_i$ allows us to construct a Pauli-string of length $n$ using $2(n-1)$ iSWAPs. Since the Pauli operators are full rank, one can drastically reduce the depth of the circuit from $\mathcal{O}(n)$ to $\mathcal{O}(\text{log}(n))$ by using Eq.~\eqref{vc}, with $\prod_{i=1}^{n}S_i=U_{\text{tot}}^\dag R_1 U_{\text{tot}}$, without needing any ancillary qubits.

Note further that instead of increasing the length of the string using  $U_{i,j}$, one can also use the inverse operation to remove a qubit from the string, e.g.~to form a disconnected string from a nearest neighbour architecture.

\section{Number- and parity-conserving strings}
We now turn to generating many-body operators that act conditionally only within the fixed excitation-number subspace. These are natural fit for a two-body, exchange gate interaction, $G^+_{i,j}=\sigma_i^+\sigma_j^-+\sigma_i^-\sigma_j^+$, which has the same symmetry, noting also its rank is smaller than $\dim(\mathcal H_i\otimes \mathcal H_j)$. Defining $F^-_{i,j}=-i(\sigma_i^+\sigma_j^--\sigma_i^-\sigma_j^+)$, we desire strings of $R_{i,j}\in\{G_{i,j},F_{i,j}\}$. We choose $U_{i,k,l}=[Z_iG_{k,l}]^{\tfrac{\pi}{4}}$ as the (now three-qubit) entangling operation, giving $U_{i,k,l}^\dag G_{i,j}U_{i,k,l}=F_{i,j}G_{k,l}+G_{i,j}P_{\text{ker}\{G_{k,l}\}}$, where $R_{k,l}^\perp=\tfrac{1}{2}(\mathds 1+Z_kZ_l)$. Following the steps in Eq.~\eqref{h}, one obtains 
\begin{align}
\hat H=&\prod_{i=1}^{n/2}R_{2i-1,2i}+\sum_{i=1}^{n/2-1}\left(\prod_{j=1}^iR_{2j-1,2j}\right)R_{2i+1,2i+2}^\perp.
\end{align}
a many-body Hamiltonian that collectively excites and de-excites $n$ qubits in a number-conserving way.  By choosing $M_n=[Z_n]^{\tfrac{\pi}{2}}$ one can apply the decoupling sequence \eqref{dd} to pick out one particular number-conserving string  
\begin{align}
[\bar H]^\alpha = \left[\prod_{i=1}^{n/2}R_{2i-1,2i}\right]^\alpha.
\end{align}
More generally, one may desire an entangler that conserves parity, without conserving number\cite{pl}.  This can be achieved by applying local operations ($X_i$) to transform operators in the string from $G_{i,j}$ to $\sigma_i^+\sigma_j^++\sigma_i^-\sigma_j^-$.  As we detail in Appendix B, the whole sequence takes $2n-4$ $U_{i,j,k}$ gates, or equivalently $6n-10$ iSWAPs. Half as many are required if $\hat H$ is used instead. Alternatively, the many-body dynamics can be generated with the selection protocol at the cost of $n-1$ ancillas. For this, we can reuse $U_{i,j,k}=C_{i,j}X_k$ for $i,j\in q_{\text{reg}}$. This total sequence uses a total of $2n-4$ entanglers $U_{i,k,l}$ and $n+4$ Toffolis in a circuit depth of $4\log(n)+3$.


\section{Multi-control CNOT gates}
$C_{1,...,n}X_{n+1}$ gates have widespread use in quantum and reversible computation, including for circuit distillation \cite{maslov2007techniques}, unstructured search \cite{molmer2011efficient}, factorization \cite{cleve2000fast}, error-correction \cite{crow2016improved}, and linear equations system solvers \cite{aharonov2003adiabatic}.
For our constructions, let $U_{i,j,k}=[P_iX_jP_k]^{\tfrac{\pi}{2}}$  (a Toffoli gate with a relative phase \cite{maslov}, though a regular Toffoli can also be used) act on three qubits, recalling that $P_i= \ket{1}\bra{1}_i$. The chaining operation is given by $U_{i,j,k}Z_iZ_jU_{i,j,k}^\dag=-P_iZ_jZ_k+P_i^\perp Z_j$, with $P_i^\perp = \ket{0}\bra{0}_i$. Repeated application of the chaining operation on $R_{1,2}=Z_1Z_2$ following Eq.~\eqref{h} (see Appendix B) gives 
\begin{align}
\hat H_n
=&(-1)^n\prod_{i=1}^{n}P_iZ_{n+1}Z_{n+2}-\sum_{j=1}^n(-1)^{j}\prod_{k=1}^{j-1}P_kP_j^\perp Z_{j+1}.\label{Htoffoli}
\end{align}
Choosing $M=[X]^{\tfrac{\pi}{2}}$, one can use the decoupling protocol \eqref{dd} to obtain a multi-qubit-controlled rotation around an arbitrary angle.
For a phaseless multi-control CNOT gate, one can compose instead the sequence 
\begin{align}%
	C_{1,...,n}Z_{n+2}=[\hat H_{n}]^{\pi/2}M_{n+1}^\dag [\hat H_{n-1}]^{-\pi/2}M_{n+1},
	\end{align}
using $4n-2$ Toffolis, or equivalently $16n-8$ CNOT gates. This cuts  in half the size and depth of the longstanding construction proposed by \cite{barenco1,maslov1,maslov}.

A more drastic reduction results from the selection protocol, where $S_i=P_i$ and thus $\text{rank(S)}=1$. Eq.~(\ref{vc}) simplifies to
\begin{align}
C_{2,...,n+1}X_{1}=C_{\text{tot}}^\dag(C_{n+1}X_{1})C_{\text{tot}},
\end{align} 
resulting in an $n$-control CNOT gate using $2(n-1)$ Toffoli gates and $2\log (n)$ depth (cf. Fig. \ref{fig:2}). This gives exponential parallelization compared to the $O(n)$ ancilla solution found in \cite{maslov}. 


\section{Unitary coupled cluster}
Another many-body operator which is frequently used (e.g. in quantum chemistry algorithms for computing energy landscapes), is an operator which transfers population between electronic orbitals (encoded in the qubits) while maintaining electron number and spin. More generally, when the operator couples arbitrary many-qubit states, it corresponds to a sparse matrix off-diagonal element \cite{aharonov2003adiabatic}. It takes the form
\begin{align}
\text{UCC}(m,n)\equiv\prod_{i=1}^m\sigma_i^+\prod_{j=m+1}^{m+n}\sigma_j^-+h.c.\label{mn}
\end{align} 
We have seen in the section on Pauli strings how to construct $R=\prod_{i=1}^{m+n}X_i$ using $2(m+n-1)$ entanglers. Since $\prod_{i=1}^{n+m}X_i$ contains all $2^{m+n}$ combinations of products of $\sigma^+$ and $\sigma^-$, we use $U=[(\prod_{i=1}^mP_i)(\prod_{j=m+1}^nP_j^\perp)X_{n+m+1}]^{\tfrac{\pi}{2}}$, which we know how to construct from the previous section on multi-control CNOTs, computing $\hat H=U^\dag RU$ to get 
\begin{align}
	\hat H=&\prod_{i=1}^{m+n}X_i-\left(\prod_{i=1}^m\sigma_i^+\prod_{j=m+1}^{m+n}\sigma_j^-+h.c.\right)\nonumber\\&+i\left(\prod_{i=1}^m\sigma_i^+\prod_{j=m+1}^{m+n}\sigma_j^--h.c.\right)X_{m+n+1}.
	\end{align}
By setting $M=[Z_{m+n+1}]^{\tfrac{\pi}{2}}$ we can apply the decoupling protocol to construct the unitary dynamics of the $\text{UCC}(m,n)$ operator in a circuit with $m+n+1$ qubits ($q_{m+n+1}$ is an ancillary qubit) using $4(m+n-1)$ iSWAPs and $4(m+n)$ (relative phase) Toffolis,
\begin{align}
	[\bar H]^{2\alpha}=[\hat H]^{\alpha}M^\dag [\hat H]^{-\alpha}M=[\text{UCC}(m,n)]^{2\alpha}.
	\end{align}
Conventional factorization of the $\text{UCC}(m,n)$ terms into Pauli strings scales exponentially as $\mathcal O(2^{m+n-1})$ in the number of two-qubit gates cost
whereas only $36(m+n) + \mathcal O(1)$ iSWAPs are required when using our decoupling protocol. Even further decrease in composition time is once again achieved if the Pauli string and multi-CNOT gates are produced using the selection protocol, down to a depth of $O(\log n)$.


\section{Architectural considerations}  
Clearly any time and space complexity advantages will be subject to limitations set by architecture.  The presence of $O(n)$ ancillas (needed for the selection protocol) is actually fairly easy to include, as most architectures have ancillary electronic, motional or photon bus degrees of freedom. Despite much worse lifetimes typically found in these states, the (linear) tradeoff in error rate is more than made up by an exponential speedup in time and justifies their use for many-body gates, in particular when memory operations are expensive.  Note previous generic $O(\log(n))$ circuit constructions require $O(n^2)$ space \cite{moore2001parallel}, which may make implementation impractical. The adjacency graph of bodies that couple to each other in the architecture will also greatly impact composition time.  For many-body operators spanning much of the graph, the spanning tree depth will determine how many steps it will take to link distant qubits in the collective many-body operator (intermediary unneeded qubits can be removed in the subsequent step). Thus, we expect the circuit depth will scale as $O(\log(n))$ when the depth of the spanning tree is $O(\log(n))$, as in \cite{brennen2003quantum,meter2006architectural, kimble2008quantum, hall2015study,van2006distributed,beigi2011graph,thaker2006quantum}, and $O(\sqrt[d]{n})$ scaling for $d$-dimensional, nearest-neighbour architectures being expected \cite{choi2011effect,choi2012theta,michael}.  



\section{Conclusions} We have developed two protocols, the decoupling and selection compositions, to generate many-body operators in $O(n)$ time for zero or constant memory overhead, and $O(\log(n))$ time for $O(n)$ ancillas, respectively.  The former enhances previous constant-overhead approaches, with improvement ranging for prominent examples  from linear (factor of 2 for multi-control CNOTs) to exponential (UCC). Our construction to bring down further the runtime to $O(\log(n))$ depth also improves quadratically on the space requirements of previous generic methods.  Our approach is generated directly from Hamiltonian dynamics, allowing straightforward incorporation of different coupling mechanisms and architectures.

\acknowledgements{ This work was supported by the European union through ScaleQIT and by US ARL-CDQI program through cooperative agreement W911NF-15-2-0061. The authors wish to thank Pierre-Luc Dallaire-Demers, Nathan Wiebe and Ryan Babbush for helpful discussions.}

\appendix

\section{Proof for decoupling protocol}
Though the decoupling protocol could be applicable to a wide range of entangling operations, we will be looking at the specific case whereby a unitary operator $U_{i,j}$ acts on some Hermitian operator $R_i$ via
\begin{align}
	U_{i,j}^\dag R_iU_{i,j}=S_iR_j+S_i^\perp+S_iR_j^\perp,  \label{u2}
	\end{align}
stressing that the entangler's action is chosen such that it covers the example cases as well as the prerequisite of the selection protocol. We will now prove the result of the staircase circuit of Fig. 2 from the main text,
\begin{align}
	\hat H_n =& \left(\prod_{j=1}^nU_{j,j+1}\right)^\dag R_1\left(\prod_{j=1}^nU_{j,j+1}\right)\nonumber\\
    =&\left(\prod_{i=1}^nS_i\right)R_{n+1}+\sum_{m=1}^n\left(\prod_{i=1}^{m-1}S_i\right)\left(S_{m}R_{m+1}^\perp+S_m^\perp\right), \label{h2} 
	\end{align}
where we set $\hat H=\hat H_n$ to clearly indicate that Eq. $\eqref{h2}$ is the result after $n$ entanglers in the staircase. We assume Eq. \eqref{h2} holds for some fixed $n\in\mathds N$. Then, using Eq. \eqref{u2} gives
\begin{align}
	\hat H_{n+1}=&U_{n+1,n+2}^\dag \hat H_nU_{n+1,n+2}\nonumber\\=&U_{n+1,n+2}^\dag \left(\prod_{i=1}^nS_i\right)R_{n+1}U_{n+1,n+2}\nonumber\\&+\sum_{m=1}^n\left(\prod_{i=1}^{m-1}S_i\right)\left(S_{m}R_{m+1}^\perp+S_m^\perp\right)\nonumber\\=&\left(\prod_{i=1}^nS_i\right)(S_{n+1}R_{n+2}+S_{n+1}^\perp+R_{n+1}R_{n+2}^\perp)\nonumber\\&+\sum_{m=1}^n\left(\prod_{i=1}^{m-1}S_i\right)\left(S_{m}R_{m+1}^\perp+S_m^\perp\right)\nonumber\\
	=&\left(\prod_{i=1}^{n+1}S_i\right)R_{n+2}+\sum_{m=1}^{n+1}\left(\prod_{i=1}^{m-1}S_i\right)\left(S_m^\perp+S_{m}R_{m+1}^\perp\right),
	\end{align}
therefore, Eq. \eqref{h2} must hold for all $n\in\mathds N$. 

\section{Gate derivations}

In the following examples we will use the modified Euler identity 
\begin{align}
[A_i]^\alpha= P_{\text{ker}\{A_i\}}+\cos(\alpha)P_{\text{supp}\{A_i\}}-i\sin(\alpha)A_i,\label{a13}
\end{align}
\begin{widetext}
where $A_i$ is a Hermitian operator that squares to the identity on its support, i.e. $A_i^2=P_{\text{supp}\{A_i\}}$, and projector $P_{\text{ker}\{A_i\}}$ is the identity operator on the kernel of $A_i$.
Moving to tensor products of such operators, e.g. $S_i$ and $B_j$, we have 
\begin{align}
[S_iB_j]^\alpha R_i [S_iB_j]^{-\alpha}=&P_{\text{ker}\{B_j\}}R_i+P_{\text{supp}\{B_j\}} P_{\text{ker}\{S_i\}}R_iP_{\text{ker}\{S_i\}} +\cos(\alpha)P_{\text{supp}\{B_j\}} P_{\text{ker}\{S_i\}}R_iP_{\text{supp}\{S_i\}}\nonumber\\&+\cos(\alpha)P_{\text{supp}\{B_j\}}P_{\text{supp}\{S_i\}}R_iP_{\text{ker}\{S_i\}}
+\cos^2(\alpha)P_{\text{supp}\{B_j\}} P_{\text{supp}\{S_i\}}R_iP_{\text{supp}\{S_i\}}\nonumber\\ &+\sin^2(\alpha)P_{\text{supp}\{B_j\}} S_iR_iS_i +i\sin(\alpha)B_j(S_iR_iP_{\text{ker}\{S_i\}}-P_{\text{ker}\{S_i\}} R_iS_i)\nonumber\\
&-i\cos(\alpha)\sin(\alpha)B_j(S_iR_iP_{\text{supp}\{S_i\}}-P_{\text{supp}\{S_i\}}R_iS_i)\label{trafo}.
\end{align}
\end{widetext}
\subsection{Pauli string generation by XX gates}
For  M\o{}lmer-S\o{}rensen composition \cite{molmer} we have $S_i=X_i$ and $B_j=X_j$ and since Pauli-operators have full rank, $P_{\text{supp}\{S_i\}}=\mathds 1_i$ and $P_{\text{supp}\{B_j\}}=\mathds 1_j$. Setting $R_i=Y_i$ and $\alpha=\pi/4$, Eq. \eqref{trafo} reduces to 
\begin{align}
	[S_iB_j]^{-\alpha} R_i [S_iB_j]^{\alpha}=& \cos^2(\alpha)R_i+\sin^2(\alpha) S_iR_iS_i\nonumber\\
    &+i\cos(\alpha)\sin(\alpha)B_j[S_i,R_i],
	\end{align}
which, using the group properties of Pauli operators $[X,Y]=2iZ$, gives
\begin{align}
[X_iX_j]^{-\tfrac{\pi}{4}} Y_i [X_iX_j]^{\tfrac{\pi}{4}}= &\frac{1}{2}Y_i+\frac{1}{2} X_iY_iX_i\nonumber\\
    &+i\cos(\alpha)\sin(\alpha)(X_iY_i-Y_iX_i)X_j\nonumber\\
    =&-Z_iX_j
\end{align}

\subsection{Pauli string generation by CNOT gates}
For CNOT gates, one can increment the string length by one by sandwiching a single Pauli-$Z$ operator in between two CNOTs \cite{nielsen},
\begin{align}
CNOT_{i,j}Z_iCNOT_{i,j}=&(\ket{0}\bra{0}_i\mathds 1_j+\ket{1}\bra{1}_i X_j)Z_i\nonumber\\
    &\times(\ket{0}\bra{0}_i\mathds 1_j+\ket{1}\bra{1}_i X_j)\nonumber\\
    =&(\ket{0}\bra{0}_i Z_j-\ket{1}\bra{1}_i Z_j)\nonumber\\
    =&Z_1Z_2.
\end{align}

\subsection{Pauli string generation by iSWAP gates}

We now consider the flip-flop interaction as the generator, where $U_{i,k}=[\sigma_i^+\sigma_j^-+\sigma_i^-\sigma_j^+]^\alpha$. For an iSWAP operation we have $\alpha=\pi/2$, and if acting on $R_i=X_i$ we get 
\begin{align}
	\text{iSWAP}_{i,j}^\dag Y_i\text{iSWAP}_{i,j}=&\frac{1}{4}(\mathds 1+Z_iZ_j+iX_iX_j+iY_iY_j)Y_i\nonumber\\
    &\times(\mathds 1+Z_iZ_j-iX_iX_j-iY_iY_j)\nonumber\\
    =&-Z_iX_j.
	\end{align}

\subsection{Number conserving strings by iSWAP gates}
For composing number-conserving strings we have $U_{i,k,l}=[Z_iG_{kl}]^{\tfrac{\pi}{4}}$,
with $G_{i,j}=\sigma_i^+\sigma_j^-+\sigma_i^-\sigma_j^+$, $F_{i,j}=-i(\sigma_i^+\sigma_j^--\sigma_i^-\sigma_j^+)$,  $P_{\text{ker}\{G_{k,l}\}}=\frac{1}{2}(\mathds 1+Z_kZ_l)$ and $P_{\text{supp}\{G_{k,l}\}}=\frac{1}{2}(\mathds 1-Z_kZ_l)$. Applying Eq. \eqref{trafo} with $S=Z_i$,$B=G_{k,l}$ and $R=G_{i,j}$ we get 
\begin{align}
	[Z_iG_{kl}]^{-\tfrac{\pi}{4}}G_{i,j}[Z_iG_{kl}]^{\tfrac{\pi}{4}}=&G_{i,j}P_{\text{ker}\{G_{k,l}\}}+\frac{1}{2}G_{i,j}P_{\text{supp}\{G_{k,l}\}}\nonumber\\
    &+\frac{1}{2}Z_iG_{i,j}Z_iP_{\text{supp}\{G_{k,l}\}}\nonumber\\
    &+i\frac{1}{2}(Z_iG_{i,j}-G_{i,j}Z_i)G_{k,l}\nonumber\\
	=&G_{i,j}P_{\text{ker}\{G_{k,l}\}}+F_{i,j}G_{k,l},
	\end{align}
since $Z_iG_{i,j}=-iF_{i,j}$.
One can similarly show that $U_{i,k,l}F_{i,j}U_{i,k,l}^\dagger=F_{i,j}P_{G_{k,l}^\perp}+G_{i,j}G_{k,l}$. The generator $Z_iG_{k,l}$ can be obtained from iSWAPs through $[F_{i,l}]^{-\pi/2}[G_{i,k}]^\alpha[F_{i,l}]^{\pi/2}=[Z_iG_{k,l}]^\alpha$.



For simplicity, we let $n$ be even. We have
\begin{align}
	U_{1,3,4}^\dag G_{1,2}U_{1,3,4}=F_{1,2}G_{3,4}+G_{1,2}P_{\text{ker}\{G_{3,4}\}}.\label{as}
	\end{align} 
Note that $P_{\text{ker}\{G_{3,4}\}}$ commutes with the next entangler of the staircase circuit of Fig. 2 in the main letter, yielding
\begin{align}
	U_{3,5,6}^\dag U_{1,3,4}^\dag G_{1,2}U_{1,3,4} U_{3,5,6}=&F_{1,2}F_{3,4}G_{5,6}+G_{1,2}P_{\text{ker}\{G_{3,4}\}}\nonumber\\
    &+F_{1,2}G_{3,4}P_{\text{ker}\{G_{5,6}\}}.
	\end{align}
Clearly, we after the $n/2$-th entangler we get
\begin{align}
	\hat H_{\frac{n}{2}}=&\left(\prod_{i=1}^{n/2} U_{2i,2i+1,2i+2}\right)^\dag G_{1,2}\left(\prod_{i=1}^{n/2} U_{2i,2i+1,2i+2}\right)\nonumber\\
	=&G_{1,2}P_{\text{ker}\{G_{3,4}\}}+\left(\prod_{m=1}^{n/2-1}F_{2m-1,2m}\right)G_{n-1,n}\nonumber\\+&\sum_{m=1}^{n/2-2}\left(\prod_{k=1}^mF_{2k-1,2k}\right)G_{2m+1,2m+2}P_{\text{ker}\{G_{2m+3,2m+4}\}}\label{ass}
	\end{align}
The proof for the above equation is analogue to the one before, when we proved Eq. \eqref{h}. In terms of our more loose notation where local unitary transformations are ignored, Eq. \eqref{as} is equivalent to $U_{i,k,l}R_{i,j}U_{i,k,l}^\dag=R_{i,j}R_{k,l}+R_{i,j}R_{k,l}^\perp$ while Eq. \eqref{ass} may be written as
\begin{align}
\hat H_{\frac{n}{2}}=&\prod_{i=1}^{n/2}R_{2i-1,2i}+\sum_{i=1}^{n/2-1}\left(\prod_{j=1}^iR_{2j-1,2j}\right)R_{2i+1,2i+2}^\perp.
\end{align}

\subsection{Multi-control NOT gates via Toffoli gates}
For generating multi-control NOT gates, we have $U_{i,j,k}=[P_iX_jP_k]^{\tfrac{\pi}{2}}$ with $P_i=\ket{1}\bra{1}_i$. Note that one can just as well use a regular Toffoli for $U_{i,j,k}$ though typically the construction uses more gates than its relative phase version \cite{maslov}. Plugging into Eq. \eqref{trafo}, with $S=P_iX_j$, $B=P_k$ and $R=Z_iZ_j$ we obtain
\begin{align}
U_{i,j,k}^\dag Z_iZ_jU_{i,j,k}=&Z_iZ_jP_k^\perp + P_i^\perp Z_jP_k   +P_iZ_jP_k\nonumber\\
&=Z_iZ_jP_k^\perp +Z_jP_k\nonumber\\
&=-P_iZ_jZ_k+P_i^\perp Z_j,\nonumber\label{tt}
	\end{align}
where we used, that $P_{\text{supp}\{S\}} =P_i$, $P_{\text{ker}\{S\}} =P_i^\perp$, $P_{\text{supp}\{B\}} =P_k$, $P_{\text{ker}\{B\}} =P_k^\perp$, $P_{\text{supp}\{R\}} =\mathds 1$ and $P_{\text{ker}\{R\}} =0$. Eq.~\eqref{Htoffoli} in the main text arises as follows: In the first iteration, we have 
\begin{align}
	U_{1,2,3}^\dag Z_1Z_2U_{1,2,3}=-P_1Z_2Z_3+P_1^\perp Z_2.
	\end{align}
Since $U_{2,3,4}$ commutes with $P_1^\perp Z_2$ we get 
\begin{align}
	U_{2,3,4}^\dag U_{1,2,3}^\dag Z_1Z_2U_{1,2,3}U_{2,3,4}&=P_1^\perp Z_2-U_{2,3,4}^\dag P_1Z_2Z_3U_{2,3,4}\nonumber\\
	=&P_1P_2Z_3Z_4-P_1P_2^\perp Z_3+P_1^\perp Z_2\nonumber
	\end{align}
in the next step. Assume that for a fixed $n\in \mathds N$ the staircase circuit yields
\begin{align}
\hat H_n=&\left(\prod_{j=1}^nU_{j,j+1,j+2}\right)^\dag Z_1Z_2\left(\prod_{j=1}^nU_{j,j+1,j+2}\right)\nonumber\\=&(-1)^n\left(\prod_{i=1}^{n}P_i\right)Z_{n+1}Z_{n+2}\nonumber\\&+\sum_{j=1}^n(-1)^{j+1}\left(\prod_{k=1}^{j-1}P_k\right)P_j^\perp Z_{j+1}.\label{ttt}
\end{align}
Then, we use Eq. \eqref{tt} to compute the $n+1$-th step,
\begin{align}
	\hat H_{n+1}=&U_{n+1,n+2,n+3}^\dag  \hat H_{n}U_{n+1,n+2,n+3}\nonumber\\
	=&(-1)^n\left(\prod_{i=1}^{n}P_i\right)(-P_{n+1}Z_{n+2}Z_{n+3}+P_{n+1}^\perp Z_{n+2})\nonumber\\
    &+\sum_{j=1}^n(-1)^{j+1}\left(\prod_{k=1}^{j-1}P_k\right)P_j^\perp Z_{j+1}\nonumber\\
	=&(-1)^{n+1}\left(\prod_{i=1}^{n+1}P_i\right)Z_{n+2}Z_{n+3}\nonumber\\
    &+\sum_{j=1}^{n+1}(-1)^{j+1}\left(\prod_{k=1}^{j-1}P_k\right)P_j^\perp Z_{j+1},
	\end{align}
therefore Eq. \eqref{ttt} holds for all $n\in \mathds N$. At this point either the usual mirroring pulse can be used to decouple unwanted terms, or, as mentioned in the main text, a full rotation on the last qubit \cite{barenco1} (here, the qubit with the highest index) can be used to remove the unwanted phase and generate a phase-less multi-control Toffoli.


\begin{thebibliography}{59}%
\makeatletter
\providecommand \@ifxundefined [1]{%
 \@ifx{#1\undefined}
}%
\providecommand \@ifnum [1]{%
 \ifnum #1\expandafter \@firstoftwo
 \else \expandafter \@secondoftwo
 \fi
}%
\providecommand \@ifx [1]{%
 \ifx #1\expandafter \@firstoftwo
 \else \expandafter \@secondoftwo
 \fi
}%
\providecommand \natexlab [1]{#1}%
\providecommand \enquote  [1]{``#1''}%
\providecommand \bibnamefont  [1]{#1}%
\providecommand \bibfnamefont [1]{#1}%
\providecommand \citenamefont [1]{#1}%
\providecommand \href@noop [0]{\@secondoftwo}%
\providecommand \@href[1]{\@@startlink{#1}\@@href}%
\providecommand \@@href[1]{\endgroup#1\@@endlink}%
\providecommand \@sanitize@url [0]{\catcode `\\12\catcode `\$12\catcode
  `\&12\catcode `\#12\catcode `\^12\catcode `\_12\catcode `\%12\relax}%
\providecommand \@@startlink[1]{}%
\providecommand \@@endlink[0]{}%
\providecommand \Eprint [0]{\href }%
\providecommand \doibase [0]{http://dx.doi.org/}%
\providecommand \selectlanguage [0]{\@gobble}%
\providecommand \bibinfo  [0]{\@secondoftwo}%
\providecommand \bibfield  [0]{\@secondoftwo}%
\providecommand \translation [1]{[#1]}%
\providecommand \BibitemOpen [0]{}%
\providecommand \bibitemStop [0]{}%
\providecommand \bibitemNoStop [0]{.\EOS\space}%
\providecommand \EOS [0]{\spacefactor3000\relax}%
\providecommand \BibitemShut  [1]{\csname bibitem#1\endcsname}%
\let\auto@bib@innerbib\@empty
\bibitem [{\citenamefont {Cleve}\ and\ \citenamefont
  {Watrous}(2000)}]{cleve2000fast}%
  \BibitemOpen
  \bibfield  {author} {\bibinfo {author} {\bibfnamefont {R.}~\bibnamefont
  {Cleve}}\ and\ \bibinfo {author} {\bibfnamefont {J.}~\bibnamefont
  {Watrous}},\ }in\ \href@noop {} {\emph {\bibinfo {booktitle} {Foundations of
  Computer Science, 2000. Proceedings. 41st Annual Symposium on}}}\ (\bibinfo
  {organization} {IEEE},\ \bibinfo {year} {2000})\ pp.\ \bibinfo {pages}
  {526--536}\BibitemShut {NoStop}%
\bibitem [{\citenamefont {Shor}(1999)}]{shor}%
  \BibitemOpen
  \bibfield  {author} {\bibinfo {author} {\bibfnamefont {P.~W.}\ \bibnamefont
  {Shor}},\ }\href@noop {} {\bibfield  {journal} {\bibinfo  {journal} {SIAM
  review}\ }\textbf {\bibinfo {volume} {41}},\ \bibinfo {pages} {303} (\bibinfo
  {year} {1999})}\BibitemShut {NoStop}%
\bibitem [{\citenamefont {Lloyd}(1996)}]{lloyd0}%
  \BibitemOpen
  \bibfield  {author} {\bibinfo {author} {\bibfnamefont {S.}~\bibnamefont
  {Lloyd}},\ }\href@noop {} {\bibfield  {journal} {\bibinfo  {journal}
  {Science}\ }\textbf {\bibinfo {volume} {273}},\ \bibinfo {pages} {1073}
  (\bibinfo {year} {1996})}\BibitemShut {NoStop}%
\bibitem [{\citenamefont {Harrow}\ \emph {et~al.}(2009)\citenamefont {Harrow},
  \citenamefont {Hassidim},\ and\ \citenamefont {Lloyd}}]{harrow2009quantum}%
  \BibitemOpen
  \bibfield  {author} {\bibinfo {author} {\bibfnamefont {A.~W.}\ \bibnamefont
  {Harrow}}, \bibinfo {author} {\bibfnamefont {A.}~\bibnamefont {Hassidim}}, \
  and\ \bibinfo {author} {\bibfnamefont {S.}~\bibnamefont {Lloyd}},\
  }\href@noop {} {\bibfield  {journal} {\bibinfo  {journal} {Physical review
  letters}\ }\textbf {\bibinfo {volume} {103}},\ \bibinfo {pages} {150502}
  (\bibinfo {year} {2009})}\BibitemShut {NoStop}%
\bibitem [{\citenamefont {Aspuru-Guzik}\ \emph {et~al.}(2005)\citenamefont
  {Aspuru-Guzik}, \citenamefont {Dutoi}, \citenamefont {Love},\ and\
  \citenamefont {Head-Gordon}}]{aag1}%
  \BibitemOpen
  \bibfield  {author} {\bibinfo {author} {\bibfnamefont {A.}~\bibnamefont
  {Aspuru-Guzik}}, \bibinfo {author} {\bibfnamefont {A.~D.}\ \bibnamefont
  {Dutoi}}, \bibinfo {author} {\bibfnamefont {P.~J.}\ \bibnamefont {Love}}, \
  and\ \bibinfo {author} {\bibfnamefont {M.}~\bibnamefont {Head-Gordon}},\
  }\href@noop {} {\bibfield  {journal} {\bibinfo  {journal} {Science}\ }\textbf
  {\bibinfo {volume} {309}},\ \bibinfo {pages} {1704} (\bibinfo {year}
  {2005})}\BibitemShut {NoStop}%
\bibitem [{\citenamefont {Barends}\ \emph {et~al.}(2015)\citenamefont
  {Barends}, \citenamefont {Lamata}, \citenamefont {Kelly}, \citenamefont
  {Garc{\'\i}a-{\'A}lvarez}, \citenamefont {Fowler}, \citenamefont {Megrant},
  \citenamefont {Jeffrey}, \citenamefont {White}, \citenamefont {Sank},
  \citenamefont {Mutus} \emph {et~al.}}]{barends1}%
  \BibitemOpen
  \bibfield  {author} {\bibinfo {author} {\bibfnamefont {R.}~\bibnamefont
  {Barends}}, \bibinfo {author} {\bibfnamefont {L.}~\bibnamefont {Lamata}},
  \bibinfo {author} {\bibfnamefont {J.}~\bibnamefont {Kelly}}, \bibinfo
  {author} {\bibfnamefont {L.}~\bibnamefont {Garc{\'\i}a-{\'A}lvarez}},
  \bibinfo {author} {\bibfnamefont {A.}~\bibnamefont {Fowler}}, \bibinfo
  {author} {\bibfnamefont {A.}~\bibnamefont {Megrant}}, \bibinfo {author}
  {\bibfnamefont {E.}~\bibnamefont {Jeffrey}}, \bibinfo {author} {\bibfnamefont
  {T.}~\bibnamefont {White}}, \bibinfo {author} {\bibfnamefont
  {D.}~\bibnamefont {Sank}}, \bibinfo {author} {\bibfnamefont {J.}~\bibnamefont
  {Mutus}},  \emph {et~al.},\ }\href@noop {} {\bibfield  {journal} {\bibinfo
  {journal} {Nature communications}\ }\textbf {\bibinfo {volume} {6}} (\bibinfo
  {year} {2015})}\BibitemShut {NoStop}%
\bibitem [{\citenamefont {Whitfield}\ \emph {et~al.}(2011)\citenamefont
  {Whitfield}, \citenamefont {Biamonte},\ and\ \citenamefont
  {Aspuru-Guzik}}]{aag2}%
  \BibitemOpen
  \bibfield  {author} {\bibinfo {author} {\bibfnamefont {J.~D.}\ \bibnamefont
  {Whitfield}}, \bibinfo {author} {\bibfnamefont {J.}~\bibnamefont {Biamonte}},
  \ and\ \bibinfo {author} {\bibfnamefont {A.}~\bibnamefont {Aspuru-Guzik}},\
  }\href@noop {} {\bibfield  {journal} {\bibinfo  {journal} {Molecular
  Physics}\ }\textbf {\bibinfo {volume} {109}},\ \bibinfo {pages} {735}
  (\bibinfo {year} {2011})}\BibitemShut {NoStop}%
\bibitem [{\citenamefont {Kassal}\ \emph {et~al.}(2011)\citenamefont {Kassal},
  \citenamefont {Whitfield}, \citenamefont {Perdomo-Ortiz}, \citenamefont
  {Yung},\ and\ \citenamefont {Aspuru-Guzik}}]{aag3}%
  \BibitemOpen
  \bibfield  {author} {\bibinfo {author} {\bibfnamefont {I.}~\bibnamefont
  {Kassal}}, \bibinfo {author} {\bibfnamefont {J.~D.}\ \bibnamefont
  {Whitfield}}, \bibinfo {author} {\bibfnamefont {A.}~\bibnamefont
  {Perdomo-Ortiz}}, \bibinfo {author} {\bibfnamefont {M.-H.}\ \bibnamefont
  {Yung}}, \ and\ \bibinfo {author} {\bibfnamefont {A.}~\bibnamefont
  {Aspuru-Guzik}},\ }\href@noop {} {\bibfield  {journal} {\bibinfo  {journal}
  {Annual review of physical chemistry}\ }\textbf {\bibinfo {volume} {62}},\
  \bibinfo {pages} {185} (\bibinfo {year} {2011})}\BibitemShut {NoStop}%
\bibitem [{\citenamefont {Wecker}\ \emph {et~al.}(2014)\citenamefont {Wecker},
  \citenamefont {Bauer}, \citenamefont {Clark}, \citenamefont {Hastings},\ and\
  \citenamefont {Troyer}}]{troyer}%
  \BibitemOpen
  \bibfield  {author} {\bibinfo {author} {\bibfnamefont {D.}~\bibnamefont
  {Wecker}}, \bibinfo {author} {\bibfnamefont {B.}~\bibnamefont {Bauer}},
  \bibinfo {author} {\bibfnamefont {B.~K.}\ \bibnamefont {Clark}}, \bibinfo
  {author} {\bibfnamefont {M.~B.}\ \bibnamefont {Hastings}}, \ and\ \bibinfo
  {author} {\bibfnamefont {M.}~\bibnamefont {Troyer}},\ }\href@noop {}
  {\bibfield  {journal} {\bibinfo  {journal} {Physical Review A}\ }\textbf
  {\bibinfo {volume} {90}},\ \bibinfo {pages} {022305} (\bibinfo {year}
  {2014})}\BibitemShut {NoStop}%
\bibitem [{\citenamefont {Babbush}\ \emph {et~al.}(2013)\citenamefont
  {Babbush}, \citenamefont {Love},\ and\ \citenamefont {Aspuru-Guzik}}]{ryan1}%
  \BibitemOpen
  \bibfield  {author} {\bibinfo {author} {\bibfnamefont {R.}~\bibnamefont
  {Babbush}}, \bibinfo {author} {\bibfnamefont {P.~J.}\ \bibnamefont {Love}}, \
  and\ \bibinfo {author} {\bibfnamefont {A.}~\bibnamefont {Aspuru-Guzik}},\
  }\href@noop {} {\bibfield  {journal} {\bibinfo  {journal} {arXiv preprint
  arXiv:1311.3967}\ } (\bibinfo {year} {2013})}\BibitemShut {NoStop}%
\bibitem [{\citenamefont {Wiebe}\ \emph {et~al.}(2011)\citenamefont {Wiebe},
  \citenamefont {Berry}, \citenamefont {H{\o}yer},\ and\ \citenamefont
  {Sanders}}]{wiebe2011simulating}%
  \BibitemOpen
  \bibfield  {author} {\bibinfo {author} {\bibfnamefont {N.}~\bibnamefont
  {Wiebe}}, \bibinfo {author} {\bibfnamefont {D.~W.}\ \bibnamefont {Berry}},
  \bibinfo {author} {\bibfnamefont {P.}~\bibnamefont {H{\o}yer}}, \ and\
  \bibinfo {author} {\bibfnamefont {B.~C.}\ \bibnamefont {Sanders}},\
  }\href@noop {} {\bibfield  {journal} {\bibinfo  {journal} {Journal of Physics
  A: Mathematical and Theoretical}\ }\textbf {\bibinfo {volume} {44}},\
  \bibinfo {pages} {445308} (\bibinfo {year} {2011})}\BibitemShut {NoStop}%
\bibitem [{\citenamefont {Peruzzo}\ \emph {et~al.}(2013)\citenamefont
  {Peruzzo}, \citenamefont {McClean}, \citenamefont {Shadbolt}, \citenamefont
  {Yung}, \citenamefont {Zhou}, \citenamefont {Love}, \citenamefont
  {Aspuru-Guzik},\ and\ \citenamefont {O'brien}}]{jarrod1}%
  \BibitemOpen
  \bibfield  {author} {\bibinfo {author} {\bibfnamefont {A.}~\bibnamefont
  {Peruzzo}}, \bibinfo {author} {\bibfnamefont {J.}~\bibnamefont {McClean}},
  \bibinfo {author} {\bibfnamefont {P.}~\bibnamefont {Shadbolt}}, \bibinfo
  {author} {\bibfnamefont {M.-H.}\ \bibnamefont {Yung}}, \bibinfo {author}
  {\bibfnamefont {X.-Q.}\ \bibnamefont {Zhou}}, \bibinfo {author}
  {\bibfnamefont {P.~J.}\ \bibnamefont {Love}}, \bibinfo {author}
  {\bibfnamefont {A.}~\bibnamefont {Aspuru-Guzik}}, \ and\ \bibinfo {author}
  {\bibfnamefont {J.~L.}\ \bibnamefont {O'brien}},\ }\href@noop {} {\bibfield
  {journal} {\bibinfo  {journal} {arXiv preprint arXiv:1304.3061}\ } (\bibinfo
  {year} {2013})}\BibitemShut {NoStop}%
\bibitem [{\citenamefont {Dallaire-Demers}\ and\ \citenamefont
  {Wilhelm}(2016)}]{pl}%
  \BibitemOpen
  \bibfield  {author} {\bibinfo {author} {\bibfnamefont {P.-L.}\ \bibnamefont
  {Dallaire-Demers}}\ and\ \bibinfo {author} {\bibfnamefont {F.~K.}\
  \bibnamefont {Wilhelm}},\ }\href@noop {} {\bibfield  {journal} {\bibinfo
  {journal} {Physical Review A}\ }\textbf {\bibinfo {volume} {93}},\ \bibinfo
  {pages} {032303} (\bibinfo {year} {2016})}\BibitemShut {NoStop}%
\bibitem [{\citenamefont {Toloui}\ and\ \citenamefont {Love}(2013)}]{borzu}%
  \BibitemOpen
  \bibfield  {author} {\bibinfo {author} {\bibfnamefont {B.}~\bibnamefont
  {Toloui}}\ and\ \bibinfo {author} {\bibfnamefont {P.~J.}\ \bibnamefont
  {Love}},\ }\href@noop {} {\bibfield  {journal} {\bibinfo  {journal} {arXiv
  preprint arXiv:1312.2579}\ } (\bibinfo {year} {2013})}\BibitemShut {NoStop}%
\bibitem [{\citenamefont {Babbush}\ \emph {et~al.}(2016)\citenamefont
  {Babbush}, \citenamefont {Berry}, \citenamefont {Kivlichan}, \citenamefont
  {Wei}, \citenamefont {Love},\ and\ \citenamefont {Aspuru-Guzik}}]{ryan2}%
  \BibitemOpen
  \bibfield  {author} {\bibinfo {author} {\bibfnamefont {R.}~\bibnamefont
  {Babbush}}, \bibinfo {author} {\bibfnamefont {D.~W.}\ \bibnamefont {Berry}},
  \bibinfo {author} {\bibfnamefont {I.~D.}\ \bibnamefont {Kivlichan}}, \bibinfo
  {author} {\bibfnamefont {A.~Y.}\ \bibnamefont {Wei}}, \bibinfo {author}
  {\bibfnamefont {P.~J.}\ \bibnamefont {Love}}, \ and\ \bibinfo {author}
  {\bibfnamefont {A.}~\bibnamefont {Aspuru-Guzik}},\ }\href@noop {} {\bibfield
  {journal} {\bibinfo  {journal} {New Journal of Physics}\ }\textbf {\bibinfo
  {volume} {18}},\ \bibinfo {pages} {033032} (\bibinfo {year}
  {2016})}\BibitemShut {NoStop}%
\bibitem [{\citenamefont {M{\o}lmer}\ \emph {et~al.}(2011)\citenamefont
  {M{\o}lmer}, \citenamefont {Isenhower},\ and\ \citenamefont
  {Saffman}}]{molmer2011efficient}%
  \BibitemOpen
  \bibfield  {author} {\bibinfo {author} {\bibfnamefont {K.}~\bibnamefont
  {M{\o}lmer}}, \bibinfo {author} {\bibfnamefont {L.}~\bibnamefont
  {Isenhower}}, \ and\ \bibinfo {author} {\bibfnamefont {M.}~\bibnamefont
  {Saffman}},\ }\href@noop {} {\bibfield  {journal} {\bibinfo  {journal}
  {Journal of Physics B: Atomic, Molecular and Optical Physics}\ }\textbf
  {\bibinfo {volume} {44}},\ \bibinfo {pages} {184016} (\bibinfo {year}
  {2011})}\BibitemShut {NoStop}%
\bibitem [{\citenamefont {Page}\ \emph {et~al.}(1999)\citenamefont {Page},
  \citenamefont {Brin}, \citenamefont {Motwani},\ and\ \citenamefont
  {Winograd}}]{page}%
  \BibitemOpen
  \bibfield  {author} {\bibinfo {author} {\bibfnamefont {L.}~\bibnamefont
  {Page}}, \bibinfo {author} {\bibfnamefont {S.}~\bibnamefont {Brin}}, \bibinfo
  {author} {\bibfnamefont {R.}~\bibnamefont {Motwani}}, \ and\ \bibinfo
  {author} {\bibfnamefont {T.}~\bibnamefont {Winograd}},\ }\href@noop {} {\emph
  {\bibinfo {title} {The PageRank citation ranking: Bringing order to the
  web.}}},\ \bibinfo {type} {Tech. Rep.}\ (\bibinfo  {institution} {Stanford
  InfoLab},\ \bibinfo {year} {1999})\BibitemShut {NoStop}%
\bibitem [{\citenamefont {Crow}\ \emph {et~al.}(2016)\citenamefont {Crow},
  \citenamefont {Joynt},\ and\ \citenamefont {Saffman}}]{crow2016improved}%
  \BibitemOpen
  \bibfield  {author} {\bibinfo {author} {\bibfnamefont {D.}~\bibnamefont
  {Crow}}, \bibinfo {author} {\bibfnamefont {R.}~\bibnamefont {Joynt}}, \ and\
  \bibinfo {author} {\bibfnamefont {M.}~\bibnamefont {Saffman}},\ }\href@noop
  {} {\bibfield  {journal} {\bibinfo  {journal} {Physical Review Letters}\
  }\textbf {\bibinfo {volume} {117}},\ \bibinfo {pages} {130503} (\bibinfo
  {year} {2016})}\BibitemShut {NoStop}%
\bibitem [{\citenamefont {Berry}(2014)}]{berry}%
  \BibitemOpen
  \bibfield  {author} {\bibinfo {author} {\bibfnamefont {D.~W.}\ \bibnamefont
  {Berry}},\ }\href@noop {} {\bibfield  {journal} {\bibinfo  {journal} {Journal
  of Physics A: Mathematical and Theoretical}\ }\textbf {\bibinfo {volume}
  {47}},\ \bibinfo {pages} {105301} (\bibinfo {year} {2014})}\BibitemShut
  {NoStop}%
\bibitem [{\citenamefont {Aharonov}\ and\ \citenamefont
  {Ta-Shma}(2003)}]{aharonov2003adiabatic}%
  \BibitemOpen
  \bibfield  {author} {\bibinfo {author} {\bibfnamefont {D.}~\bibnamefont
  {Aharonov}}\ and\ \bibinfo {author} {\bibfnamefont {A.}~\bibnamefont
  {Ta-Shma}},\ }in\ \href@noop {} {\emph {\bibinfo {booktitle} {Proceedings of
  the thirty-fifth annual ACM symposium on Theory of computing}}}\ (\bibinfo
  {organization} {ACM},\ \bibinfo {year} {2003})\ pp.\ \bibinfo {pages}
  {20--29}\BibitemShut {NoStop}%
\bibitem [{\citenamefont {M{\o}lmer}\ and\ \citenamefont
  {S{\o}rensen}(1999)}]{molmer}%
  \BibitemOpen
  \bibfield  {author} {\bibinfo {author} {\bibfnamefont {K.}~\bibnamefont
  {M{\o}lmer}}\ and\ \bibinfo {author} {\bibfnamefont {A.}~\bibnamefont
  {S{\o}rensen}},\ }\href@noop {} {\bibfield  {journal} {\bibinfo  {journal}
  {Physical Review Letters}\ }\textbf {\bibinfo {volume} {82}},\ \bibinfo
  {pages} {1835} (\bibinfo {year} {1999})}\BibitemShut {NoStop}%
\bibitem [{\citenamefont {Blatt}(2000)}]{blatt2000quantum}%
  \BibitemOpen
  \bibfield  {author} {\bibinfo {author} {\bibfnamefont {R.}~\bibnamefont
  {Blatt}},\ }\href@noop {} {\bibfield  {journal} {\bibinfo  {journal}
  {Nature}\ }\textbf {\bibinfo {volume} {404}},\ \bibinfo {pages} {231}
  (\bibinfo {year} {2000})}\BibitemShut {NoStop}%
\bibitem [{\citenamefont {Monz}\ \emph {et~al.}(2011)\citenamefont {Monz},
  \citenamefont {Schindler}, \citenamefont {Barreiro}, \citenamefont {Chwalla},
  \citenamefont {Nigg}, \citenamefont {Coish}, \citenamefont {Harlander},
  \citenamefont {H{\"a}nsel}, \citenamefont {Hennrich},\ and\ \citenamefont
  {Blatt}}]{monz201114}%
  \BibitemOpen
  \bibfield  {author} {\bibinfo {author} {\bibfnamefont {T.}~\bibnamefont
  {Monz}}, \bibinfo {author} {\bibfnamefont {P.}~\bibnamefont {Schindler}},
  \bibinfo {author} {\bibfnamefont {J.~T.}\ \bibnamefont {Barreiro}}, \bibinfo
  {author} {\bibfnamefont {M.}~\bibnamefont {Chwalla}}, \bibinfo {author}
  {\bibfnamefont {D.}~\bibnamefont {Nigg}}, \bibinfo {author} {\bibfnamefont
  {W.~A.}\ \bibnamefont {Coish}}, \bibinfo {author} {\bibfnamefont
  {M.}~\bibnamefont {Harlander}}, \bibinfo {author} {\bibfnamefont
  {W.}~\bibnamefont {H{\"a}nsel}}, \bibinfo {author} {\bibfnamefont
  {M.}~\bibnamefont {Hennrich}}, \ and\ \bibinfo {author} {\bibfnamefont
  {R.}~\bibnamefont {Blatt}},\ }\href@noop {} {\bibfield  {journal} {\bibinfo
  {journal} {Physical Review Letters}\ }\textbf {\bibinfo {volume} {106}},\
  \bibinfo {pages} {130506} (\bibinfo {year} {2011})}\BibitemShut {NoStop}%
\bibitem [{\citenamefont {Moore}\ and\ \citenamefont
  {Nilsson}(2001)}]{moore2001parallel}%
  \BibitemOpen
  \bibfield  {author} {\bibinfo {author} {\bibfnamefont {C.}~\bibnamefont
  {Moore}}\ and\ \bibinfo {author} {\bibfnamefont {M.}~\bibnamefont
  {Nilsson}},\ }\href@noop {} {\bibfield  {journal} {\bibinfo  {journal} {SIAM
  Journal on Computing}\ }\textbf {\bibinfo {volume} {31}},\ \bibinfo {pages}
  {799} (\bibinfo {year} {2001})}\BibitemShut {NoStop}%
\bibitem [{\citenamefont {Fenner}\ \emph {et~al.}(2005)\citenamefont {Fenner},
  \citenamefont {Green}, \citenamefont {Homer},\ and\ \citenamefont
  {Zhang}}]{fenner2005bounds}%
  \BibitemOpen
  \bibfield  {author} {\bibinfo {author} {\bibfnamefont {S.}~\bibnamefont
  {Fenner}}, \bibinfo {author} {\bibfnamefont {F.}~\bibnamefont {Green}},
  \bibinfo {author} {\bibfnamefont {S.}~\bibnamefont {Homer}}, \ and\ \bibinfo
  {author} {\bibfnamefont {Y.}~\bibnamefont {Zhang}},\ }in\ \href@noop {}
  {\emph {\bibinfo {booktitle} {International Symposium on Fundamentals of
  Computation Theory}}}\ (\bibinfo {organization} {Springer},\ \bibinfo {year}
  {2005})\ pp.\ \bibinfo {pages} {44--55}\BibitemShut {NoStop}%
\bibitem [{\citenamefont {Takahashi}\ and\ \citenamefont
  {Kunihiro}(2008)}]{takahashi2008fast}%
  \BibitemOpen
  \bibfield  {author} {\bibinfo {author} {\bibfnamefont {Y.}~\bibnamefont
  {Takahashi}}\ and\ \bibinfo {author} {\bibfnamefont {N.}~\bibnamefont
  {Kunihiro}},\ }\href@noop {} {\bibfield  {journal} {\bibinfo  {journal}
  {Quantum Information \& Computation}\ }\textbf {\bibinfo {volume} {8}},\
  \bibinfo {pages} {636} (\bibinfo {year} {2008})}\BibitemShut {NoStop}%
\bibitem [{\citenamefont {Gossett}(1998)}]{gossett1998quantum}%
  \BibitemOpen
  \bibfield  {author} {\bibinfo {author} {\bibfnamefont {P.}~\bibnamefont
  {Gossett}},\ }\href@noop {} {\bibfield  {journal} {\bibinfo  {journal} {arXiv
  preprint quant-ph/9808061}\ } (\bibinfo {year} {1998})}\BibitemShut {NoStop}%
\bibitem [{\citenamefont {Draper}\ \emph {et~al.}(2004)\citenamefont {Draper},
  \citenamefont {Kutin}, \citenamefont {Rains},\ and\ \citenamefont
  {Svore}}]{svore}%
  \BibitemOpen
  \bibfield  {author} {\bibinfo {author} {\bibfnamefont {T.~G.}\ \bibnamefont
  {Draper}}, \bibinfo {author} {\bibfnamefont {S.~A.}\ \bibnamefont {Kutin}},
  \bibinfo {author} {\bibfnamefont {E.~M.}\ \bibnamefont {Rains}}, \ and\
  \bibinfo {author} {\bibfnamefont {K.~M.}\ \bibnamefont {Svore}},\ }\href@noop
  {} {\bibfield  {journal} {\bibinfo  {journal} {arXiv preprint
  quant-ph/0406142}\ } (\bibinfo {year} {2004})}\BibitemShut {NoStop}%
\bibitem [{\citenamefont {Van~Meter}\ and\ \citenamefont
  {Itoh}(2005)}]{van2005fast}%
  \BibitemOpen
  \bibfield  {author} {\bibinfo {author} {\bibfnamefont {R.}~\bibnamefont
  {Van~Meter}}\ and\ \bibinfo {author} {\bibfnamefont {K.~M.}\ \bibnamefont
  {Itoh}},\ }\href@noop {} {\bibfield  {journal} {\bibinfo  {journal} {Physical
  Review A}\ }\textbf {\bibinfo {volume} {71}},\ \bibinfo {pages} {052320}
  (\bibinfo {year} {2005})}\BibitemShut {NoStop}%
\bibitem [{\citenamefont {Maslov}\ and\ \citenamefont {Dueck}(2003)}]{maslov1}%
  \BibitemOpen
  \bibfield  {author} {\bibinfo {author} {\bibfnamefont {D.}~\bibnamefont
  {Maslov}}\ and\ \bibinfo {author} {\bibfnamefont {G.~W.}\ \bibnamefont
  {Dueck}},\ }\href@noop {} {\bibfield  {journal} {\bibinfo  {journal}
  {Electronics Letters}\ }\textbf {\bibinfo {volume} {39}},\ \bibinfo {pages}
  {1790} (\bibinfo {year} {2003})}\BibitemShut {NoStop}%
\bibitem [{\citenamefont {Barenco}\ \emph {et~al.}(1995)\citenamefont
  {Barenco}, \citenamefont {Bennett}, \citenamefont {Cleve}, \citenamefont
  {DiVincenzo}, \citenamefont {Margolus}, \citenamefont {Shor}, \citenamefont
  {Sleator}, \citenamefont {Smolin},\ and\ \citenamefont
  {Weinfurter}}]{barenco1}%
  \BibitemOpen
  \bibfield  {author} {\bibinfo {author} {\bibfnamefont {A.}~\bibnamefont
  {Barenco}}, \bibinfo {author} {\bibfnamefont {C.~H.}\ \bibnamefont
  {Bennett}}, \bibinfo {author} {\bibfnamefont {R.}~\bibnamefont {Cleve}},
  \bibinfo {author} {\bibfnamefont {D.~P.}\ \bibnamefont {DiVincenzo}},
  \bibinfo {author} {\bibfnamefont {N.}~\bibnamefont {Margolus}}, \bibinfo
  {author} {\bibfnamefont {P.}~\bibnamefont {Shor}}, \bibinfo {author}
  {\bibfnamefont {T.}~\bibnamefont {Sleator}}, \bibinfo {author} {\bibfnamefont
  {J.~A.}\ \bibnamefont {Smolin}}, \ and\ \bibinfo {author} {\bibfnamefont
  {H.}~\bibnamefont {Weinfurter}},\ }\href@noop {} {\bibfield  {journal}
  {\bibinfo  {journal} {Physical review A}\ }\textbf {\bibinfo {volume} {52}},\
  \bibinfo {pages} {3457} (\bibinfo {year} {1995})}\BibitemShut {NoStop}%
\bibitem [{\citenamefont {Maslov}(2016)}]{maslov}%
  \BibitemOpen
  \bibfield  {author} {\bibinfo {author} {\bibfnamefont {D.}~\bibnamefont
  {Maslov}},\ }\href@noop {} {\bibfield  {journal} {\bibinfo  {journal}
  {Physical Review A}\ }\textbf {\bibinfo {volume} {93}},\ \bibinfo {pages}
  {022311} (\bibinfo {year} {2016})}\BibitemShut {NoStop}%
\bibitem [{\citenamefont {Berry}\ \emph {et~al.}(2014)\citenamefont {Berry},
  \citenamefont {Childs}, \citenamefont {Cleve}, \citenamefont {Kothari},\ and\
  \citenamefont {Somma}}]{berry2014exponential}%
  \BibitemOpen
  \bibfield  {author} {\bibinfo {author} {\bibfnamefont {D.~W.}\ \bibnamefont
  {Berry}}, \bibinfo {author} {\bibfnamefont {A.~M.}\ \bibnamefont {Childs}},
  \bibinfo {author} {\bibfnamefont {R.}~\bibnamefont {Cleve}}, \bibinfo
  {author} {\bibfnamefont {R.}~\bibnamefont {Kothari}}, \ and\ \bibinfo
  {author} {\bibfnamefont {R.~D.}\ \bibnamefont {Somma}},\ }in\ \href@noop {}
  {\emph {\bibinfo {booktitle} {Proceedings of the 46th Annual ACM Symposium on
  Theory of Computing}}}\ (\bibinfo {organization} {ACM},\ \bibinfo {year}
  {2014})\ pp.\ \bibinfo {pages} {283--292}\BibitemShut {NoStop}%
\bibitem [{\citenamefont {Bartlett}\ and\ \citenamefont
  {Musia{\l}}(2007)}]{bartlett1}%
  \BibitemOpen
  \bibfield  {author} {\bibinfo {author} {\bibfnamefont {R.~J.}\ \bibnamefont
  {Bartlett}}\ and\ \bibinfo {author} {\bibfnamefont {M.}~\bibnamefont
  {Musia{\l}}},\ }\href@noop {} {\bibfield  {journal} {\bibinfo  {journal}
  {Reviews of Modern Physics}\ }\textbf {\bibinfo {volume} {79}},\ \bibinfo
  {pages} {291} (\bibinfo {year} {2007})}\BibitemShut {NoStop}%
\bibitem [{\citenamefont {Taube}\ and\ \citenamefont
  {Bartlett}(2006)}]{bartlett2}%
  \BibitemOpen
  \bibfield  {author} {\bibinfo {author} {\bibfnamefont {A.~G.}\ \bibnamefont
  {Taube}}\ and\ \bibinfo {author} {\bibfnamefont {R.~J.}\ \bibnamefont
  {Bartlett}},\ }\href@noop {} {\bibfield  {journal} {\bibinfo  {journal}
  {International journal of quantum chemistry}\ }\textbf {\bibinfo {volume}
  {106}},\ \bibinfo {pages} {3393} (\bibinfo {year} {2006})}\BibitemShut
  {NoStop}%
\bibitem [{\citenamefont {Romero}\ \emph {et~al.}(2017)\citenamefont {Romero},
  \citenamefont {Babbush}, \citenamefont {McClean}, \citenamefont {Hempel},
  \citenamefont {Love},\ and\ \citenamefont {Aspuru-Guzik}}]{jonathan}%
  \BibitemOpen
  \bibfield  {author} {\bibinfo {author} {\bibfnamefont {J.}~\bibnamefont
  {Romero}}, \bibinfo {author} {\bibfnamefont {R.}~\bibnamefont {Babbush}},
  \bibinfo {author} {\bibfnamefont {J.~R.}\ \bibnamefont {McClean}}, \bibinfo
  {author} {\bibfnamefont {C.}~\bibnamefont {Hempel}}, \bibinfo {author}
  {\bibfnamefont {P.}~\bibnamefont {Love}}, \ and\ \bibinfo {author}
  {\bibfnamefont {A.}~\bibnamefont {Aspuru-Guzik}},\ }\href@noop {} {\bibfield
  {journal} {\bibinfo  {journal} {arXiv preprint arXiv:1701.02691}\ } (\bibinfo
  {year} {2017})}\BibitemShut {NoStop}%
\bibitem [{\citenamefont {Pulay}\ and\ \citenamefont
  {Schaefer~III}(1977)}]{kutzelnigg}%
  \BibitemOpen
  \bibfield  {author} {\bibinfo {author} {\bibfnamefont {P.}~\bibnamefont
  {Pulay}}\ and\ \bibinfo {author} {\bibfnamefont {H.}~\bibnamefont
  {Schaefer~III}},\ }\href@noop {} {\bibfield  {journal} {\bibinfo  {journal}
  {Modern theoretical chemistry}\ }\textbf {\bibinfo {volume} {4}},\ \bibinfo
  {pages} {153} (\bibinfo {year} {1977})}\BibitemShut {NoStop}%
\bibitem [{\citenamefont {Hoffmann}\ and\ \citenamefont
  {Simons}(1988)}]{hoffmann}%
  \BibitemOpen
  \bibfield  {author} {\bibinfo {author} {\bibfnamefont {M.~R.}\ \bibnamefont
  {Hoffmann}}\ and\ \bibinfo {author} {\bibfnamefont {J.}~\bibnamefont
  {Simons}},\ }\href@noop {} {\bibfield  {journal} {\bibinfo  {journal} {The
  Journal of chemical physics}\ }\textbf {\bibinfo {volume} {88}},\ \bibinfo
  {pages} {993} (\bibinfo {year} {1988})}\BibitemShut {NoStop}%
\bibitem [{\citenamefont {Bera}\ \emph {et~al.}(2010)\citenamefont {Bera},
  \citenamefont {Fenner}, \citenamefont {Green},\ and\ \citenamefont
  {Homer}}]{bera}%
  \BibitemOpen
  \bibfield  {author} {\bibinfo {author} {\bibfnamefont {D.}~\bibnamefont
  {Bera}}, \bibinfo {author} {\bibfnamefont {S.}~\bibnamefont {Fenner}},
  \bibinfo {author} {\bibfnamefont {F.}~\bibnamefont {Green}}, \ and\ \bibinfo
  {author} {\bibfnamefont {S.}~\bibnamefont {Homer}},\ }\href@noop {}
  {\bibfield  {journal} {\bibinfo  {journal} {Quantum Information \&
  Computation}\ }\textbf {\bibinfo {volume} {10}},\ \bibinfo {pages} {16}
  (\bibinfo {year} {2010})}\BibitemShut {NoStop}%
\bibitem [{mik()}]{mikecomment}%
  \BibitemOpen
  \href@noop {}  {\bibinfo {title} {A sufficient condition that such a
  transformation exists is that the eigenvalues $\{\lambda_{n+1}^{(i)}\}$ of
  $r_{n+1}$ are symmetric, i.e. for each eigenvalue $\lambda_{n+1}^{(i)}$ there
  exists a corresponding eigenvalue $\lambda_{n+1}^{(j)}$ ($i\neq j$) such that
  $\lambda_{n+1}^{(i)}=-\lambda_{n+1}^{(j)}$}}\BibitemShut {NoStop}%
\bibitem [{\citenamefont {Nielsen}\ and\ \citenamefont
  {Chuang}(2002)}]{nielsen}%
  \BibitemOpen
  \bibfield  {author} {\bibinfo {author} {\bibfnamefont {M.~A.}\ \bibnamefont
  {Nielsen}}\ and\ \bibinfo {author} {\bibfnamefont {I.}~\bibnamefont
  {Chuang}},\ }\href@noop {} {\enquote {\bibinfo {title} {Quantum computation
  and quantum information},}\ } (\bibinfo {year} {2002})\BibitemShut {NoStop}%
\bibitem [{\citenamefont {Casanova}\ \emph {et~al.}(2012)\citenamefont
  {Casanova}, \citenamefont {Mezzacapo}, \citenamefont {Lamata},\ and\
  \citenamefont {Solano}}]{mezzacapo}%
  \BibitemOpen
  \bibfield  {author} {\bibinfo {author} {\bibfnamefont {J.}~\bibnamefont
  {Casanova}}, \bibinfo {author} {\bibfnamefont {A.}~\bibnamefont {Mezzacapo}},
  \bibinfo {author} {\bibfnamefont {L.}~\bibnamefont {Lamata}}, \ and\ \bibinfo
  {author} {\bibfnamefont {E.}~\bibnamefont {Solano}},\ }\href@noop {}
  {\bibfield  {journal} {\bibinfo  {journal} {Physical review letters}\
  }\textbf {\bibinfo {volume} {108}},\ \bibinfo {pages} {190502} (\bibinfo
  {year} {2012})}\BibitemShut {NoStop}%
\bibitem [{\citenamefont {S{\o}rensen}\ and\ \citenamefont
  {M{\o}lmer}(2000)}]{molmerII}%
  \BibitemOpen
  \bibfield  {author} {\bibinfo {author} {\bibfnamefont {A.}~\bibnamefont
  {S{\o}rensen}}\ and\ \bibinfo {author} {\bibfnamefont {K.}~\bibnamefont
  {M{\o}lmer}},\ }\href@noop {} {\bibfield  {journal} {\bibinfo  {journal}
  {Physical Review A}\ }\textbf {\bibinfo {volume} {62}},\ \bibinfo {pages}
  {022311} (\bibinfo {year} {2000})}\BibitemShut {NoStop}%
\bibitem [{\citenamefont {Goerz}\ \emph {et~al.}(2016)\citenamefont {Goerz},
  \citenamefont {Motzoi}, \citenamefont {Whaley},\ and\ \citenamefont
  {Koch}}]{goerz}%
  \BibitemOpen
  \bibfield  {author} {\bibinfo {author} {\bibfnamefont {M.~H.}\ \bibnamefont
  {Goerz}}, \bibinfo {author} {\bibfnamefont {F.}~\bibnamefont {Motzoi}},
  \bibinfo {author} {\bibfnamefont {K.~B.}\ \bibnamefont {Whaley}}, \ and\
  \bibinfo {author} {\bibfnamefont {C.~P.}\ \bibnamefont {Koch}},\ }\href@noop
  {} {\bibfield  {journal} {\bibinfo  {journal} {
  arXiv:1606.08825}\ } (\bibinfo {year} {2016})}\BibitemShut {NoStop}%
\bibitem [{\citenamefont {{Siewert}}\ \emph {et~al.}(2000)\citenamefont
  {{Siewert}}, \citenamefont {{Fazio}}, \citenamefont {{Palma}},\ and\
  \citenamefont {{Sciacca}}}]{siewert}%
  \BibitemOpen
  \bibfield  {author} {\bibinfo {author} {\bibfnamefont {J.}~\bibnamefont
  {{Siewert}}}, \bibinfo {author} {\bibfnamefont {R.}~\bibnamefont {{Fazio}}},
  \bibinfo {author} {\bibfnamefont {G.~M.}\ \bibnamefont {{Palma}}}, \ and\
  \bibinfo {author} {\bibfnamefont {E.}~\bibnamefont {{Sciacca}}},\ }{\bibfield  {journal} {\bibinfo  {journal}
  {Journal of Low Temperature Physics}\ }\textbf {\bibinfo {volume} {118}},\
  \bibinfo {pages} {795} (\bibinfo {year} {2000})}\BibitemShut {NoStop}%
\bibitem [{\citenamefont {Imamog}\ \emph {et~al.}(1999)\citenamefont {Imamog},
  \citenamefont {Awschalom}, \citenamefont {Burkard}, \citenamefont
  {DiVincenzo}, \citenamefont {Loss}, \citenamefont {Sherwin}, \citenamefont
  {Small} \emph {et~al.}}]{imamoglo}%
  \BibitemOpen
  \bibfield  {author} {\bibinfo {author} {\bibfnamefont {A.}~\bibnamefont
  {Imamog}}, \bibinfo {author} {\bibfnamefont {D.~D.}\ \bibnamefont
  {Awschalom}}, \bibinfo {author} {\bibfnamefont {G.}~\bibnamefont {Burkard}},
  \bibinfo {author} {\bibfnamefont {D.~P.}\ \bibnamefont {DiVincenzo}},
  \bibinfo {author} {\bibfnamefont {D.}~\bibnamefont {Loss}}, \bibinfo {author}
  {\bibfnamefont {M.}~\bibnamefont {Sherwin}}, \bibinfo {author} {\bibfnamefont
  {A.}~\bibnamefont {Small}},  \emph {et~al.},\ }\href@noop {} {\bibfield
  {journal} {\bibinfo  {journal} {Physical Review Letters}\ }\textbf {\bibinfo
  {volume} {83}},\ \bibinfo {pages} {4204} (\bibinfo {year}
  {1999})}\BibitemShut {NoStop}%
\bibitem [{\citenamefont {Mozyrsky}\ \emph {et~al.}(2001)\citenamefont
  {Mozyrsky}, \citenamefont {Privman},\ and\ \citenamefont
  {Glasser}}]{mozyrsky}%
  \BibitemOpen
  \bibfield  {author} {\bibinfo {author} {\bibfnamefont {D.}~\bibnamefont
  {Mozyrsky}}, \bibinfo {author} {\bibfnamefont {V.}~\bibnamefont {Privman}}, \
  and\ \bibinfo {author} {\bibfnamefont {M.~L.}\ \bibnamefont {Glasser}},\
  }\href@noop {} {\bibfield  {journal} {\bibinfo  {journal} {Physical review
  letters}\ }\textbf {\bibinfo {volume} {86}},\ \bibinfo {pages} {5112}
  (\bibinfo {year} {2001})}\BibitemShut {NoStop}%
\bibitem [{\citenamefont {Maslov}\ \emph {et~al.}(2007)\citenamefont {Maslov},
  \citenamefont {Dueck},\ and\ \citenamefont {Miller}}]{maslov2007techniques}%
  \BibitemOpen
  \bibfield  {author} {\bibinfo {author} {\bibfnamefont {D.}~\bibnamefont
  {Maslov}}, \bibinfo {author} {\bibfnamefont {G.~W.}\ \bibnamefont {Dueck}}, \
  and\ \bibinfo {author} {\bibfnamefont {D.~M.}\ \bibnamefont {Miller}},\
  }\href@noop {} {\bibfield  {journal} {\bibinfo  {journal} {ACM Transactions
  on Design Automation of Electronic Systems (TODAES)}\ }\textbf {\bibinfo
  {volume} {12}},\ \bibinfo {pages} {42} (\bibinfo {year} {2007})}\BibitemShut
  {NoStop}%
\bibitem [{\citenamefont {Brennen}\ \emph {et~al.}(2003)\citenamefont
  {Brennen}, \citenamefont {Song},\ and\ \citenamefont
  {Williams}}]{brennen2003quantum}%
  \BibitemOpen
  \bibfield  {author} {\bibinfo {author} {\bibfnamefont {G.~K.}\ \bibnamefont
  {Brennen}}, \bibinfo {author} {\bibfnamefont {D.}~\bibnamefont {Song}}, \
  and\ \bibinfo {author} {\bibfnamefont {C.~J.}\ \bibnamefont {Williams}},\
  }\href@noop {} {\bibfield  {journal} {\bibinfo  {journal} {Physical Review
  A}\ }\textbf {\bibinfo {volume} {67}},\ \bibinfo {pages} {050302} (\bibinfo
  {year} {2003})}\BibitemShut {NoStop}%
\bibitem [{\citenamefont {Meter}\ and\ \citenamefont
  {Oskin}(2006)}]{meter2006architectural}%
  \BibitemOpen
  \bibfield  {author} {\bibinfo {author} {\bibfnamefont {R.~v.}\ \bibnamefont
  {Meter}}\ and\ \bibinfo {author} {\bibfnamefont {M.}~\bibnamefont {Oskin}},\
  }\href@noop {} {\bibfield  {journal} {\bibinfo  {journal} {ACM Journal on
  Emerging Technologies in Computing Systems (JETC)}\ }\textbf {\bibinfo
  {volume} {2}},\ \bibinfo {pages} {31} (\bibinfo {year} {2006})}\BibitemShut
  {NoStop}%
\bibitem [{\citenamefont {Kimble}(2008)}]{kimble2008quantum}%
  \BibitemOpen
  \bibfield  {author} {\bibinfo {author} {\bibfnamefont {H.~J.}\ \bibnamefont
  {Kimble}},\ }\href@noop {} {\bibfield  {journal} {\bibinfo  {journal}
  {Nature}\ }\textbf {\bibinfo {volume} {453}},\ \bibinfo {pages} {1023}
  (\bibinfo {year} {2008})}\BibitemShut {NoStop}%
\bibitem [{\citenamefont {Hall}\ \emph {et~al.}(2015)\citenamefont {Hall},
  \citenamefont {Novotny}, \citenamefont {Neuhaus},\ and\ \citenamefont
  {Michielsen}}]{hall2015study}%
  \BibitemOpen
  \bibfield  {author} {\bibinfo {author} {\bibfnamefont {J.}~\bibnamefont
  {Hall}}, \bibinfo {author} {\bibfnamefont {M.}~\bibnamefont {Novotny}},
  \bibinfo {author} {\bibfnamefont {T.}~\bibnamefont {Neuhaus}}, \ and\
  \bibinfo {author} {\bibfnamefont {K.}~\bibnamefont {Michielsen}},\
  }\href@noop {} {\bibfield  {journal} {\bibinfo  {journal} {Physics procedia}\
  }\textbf {\bibinfo {volume} {68}},\ \bibinfo {pages} {56} (\bibinfo {year}
  {2015})}\BibitemShut {NoStop}%
\bibitem [{\citenamefont {Van~Meter}\ \emph {et~al.}(2006)\citenamefont
  {Van~Meter}, \citenamefont {Nemoto}, \citenamefont {Munro},\ and\
  \citenamefont {Itoh}}]{van2006distributed}%
  \BibitemOpen
  \bibfield  {author} {\bibinfo {author} {\bibfnamefont {R.}~\bibnamefont
  {Van~Meter}}, \bibinfo {author} {\bibfnamefont {K.}~\bibnamefont {Nemoto}},
  \bibinfo {author} {\bibfnamefont {W.}~\bibnamefont {Munro}}, \ and\ \bibinfo
  {author} {\bibfnamefont {K.~M.}\ \bibnamefont {Itoh}},\ }\href@noop {}
  {\bibfield  {journal} {\bibinfo  {journal} {ACM SIGARCH Computer Architecture
  News}\ }\textbf {\bibinfo {volume} {34}},\ \bibinfo {pages} {354} (\bibinfo
  {year} {2006})}\BibitemShut {NoStop}%
\bibitem [{\citenamefont {Beigi}\ \emph {et~al.}(2011)\citenamefont {Beigi},
  \citenamefont {Chuang}, \citenamefont {Grassl}, \citenamefont {Shor},\ and\
  \citenamefont {Zeng}}]{beigi2011graph}%
  \BibitemOpen
  \bibfield  {author} {\bibinfo {author} {\bibfnamefont {S.}~\bibnamefont
  {Beigi}}, \bibinfo {author} {\bibfnamefont {I.}~\bibnamefont {Chuang}},
  \bibinfo {author} {\bibfnamefont {M.}~\bibnamefont {Grassl}}, \bibinfo
  {author} {\bibfnamefont {P.}~\bibnamefont {Shor}}, \ and\ \bibinfo {author}
  {\bibfnamefont {B.}~\bibnamefont {Zeng}},\ }\href@noop {} {\bibfield
  {journal} {\bibinfo  {journal} {Journal of Mathematical Physics}\ }\textbf
  {\bibinfo {volume} {52}},\ \bibinfo {pages} {022201} (\bibinfo {year}
  {2011})}\BibitemShut {NoStop}%
\bibitem [{\citenamefont {Thaker}\ \emph {et~al.}(2006)\citenamefont {Thaker},
  \citenamefont {Metodi}, \citenamefont {Cross}, \citenamefont {Chuang},\ and\
  \citenamefont {Chong}}]{thaker2006quantum}%
  \BibitemOpen
  \bibfield  {author} {\bibinfo {author} {\bibfnamefont {D.~D.}\ \bibnamefont
  {Thaker}}, \bibinfo {author} {\bibfnamefont {T.~S.}\ \bibnamefont {Metodi}},
  \bibinfo {author} {\bibfnamefont {A.~W.}\ \bibnamefont {Cross}}, \bibinfo
  {author} {\bibfnamefont {I.~L.}\ \bibnamefont {Chuang}}, \ and\ \bibinfo
  {author} {\bibfnamefont {F.~T.}\ \bibnamefont {Chong}},\ }in\ \href@noop {}
  {\emph {\bibinfo {booktitle} {ACM SIGARCH Computer Architecture News}}},\
  Vol.~\bibinfo {volume} {34}\ (\bibinfo {organization} {IEEE Computer
  Society},\ \bibinfo {year} {2006})\ pp.\ \bibinfo {pages}
  {378--390}\BibitemShut {NoStop}%
\bibitem [{\citenamefont {Choi}\ and\ \citenamefont
  {Van~Meter}(2011)}]{choi2011effect}%
  \BibitemOpen
  \bibfield  {author} {\bibinfo {author} {\bibfnamefont {B.-S.}\ \bibnamefont
  {Choi}}\ and\ \bibinfo {author} {\bibfnamefont {R.}~\bibnamefont
  {Van~Meter}},\ }\href@noop {} {\bibfield  {journal} {\bibinfo  {journal} {ACM
  Journal on Emerging Technologies in Computing Systems (JETC)}\ }\textbf
  {\bibinfo {volume} {7}},\ \bibinfo {pages} {11} (\bibinfo {year}
  {2011})}\BibitemShut {NoStop}%
\bibitem [{\citenamefont {Choi}\ and\ \citenamefont
  {Van~Meter}(2012)}]{choi2012theta}%
  \BibitemOpen
  \bibfield  {author} {\bibinfo {author} {\bibfnamefont {B.-S.}\ \bibnamefont
  {Choi}}\ and\ \bibinfo {author} {\bibfnamefont {R.}~\bibnamefont
  {Van~Meter}},\ }\href@noop {} {\bibfield  {journal} {\bibinfo  {journal} {ACM
  Journal on Emerging Technologies in Computing Systems (JETC)}\ }\textbf
  {\bibinfo {volume} {8}},\ \bibinfo {pages} {24} (\bibinfo {year}
  {2012})}\BibitemShut {NoStop}%
\bibitem [{\citenamefont {Kaicher}\ \emph {et~al.}()\citenamefont {Kaicher},
  \citenamefont {Motzoi},\ and\ \citenamefont {Wilhelm}}]{michael}%
  \BibitemOpen
  \bibfield  {author} {\bibinfo {author} {\bibfnamefont {M.}~\bibnamefont
  {Kaicher}}, \bibinfo {author} {\bibfnamefont {F.}~\bibnamefont {Motzoi}}, \
  and\ \bibinfo {author} {\bibfnamefont {F.}~\bibnamefont {Wilhelm}},\
  }\href@noop {} {}\bibinfo {note} {in preparation}\BibitemShut {NoStop}%
\end{thebibliography}
\end{document}